\documentclass[nofootinbib,twocolumn,superscriptaddress,longbibliography]{revtex4-1} 

\usepackage{latexsym}
\usepackage{graphics}
\usepackage{bm}
\usepackage[normalem]{ulem}
\usepackage{graphicx} 
\usepackage{color}
\usepackage{amsmath}
\usepackage{amssymb}
\usepackage{feynmp-auto}
\usepackage{slashed}
\usepackage{enumerate}
\usepackage{amsfonts, amssymb,amsmath,latexsym,graphics, graphicx,epsfig,multirow,comment,
feyn,slashed,xcolor,afterpage, makecell} 
\usepackage{booktabs, blindtext}
\usepackage{tabularx,afterpage}
\usepackage[colorlinks=true
,urlcolor=blue
,anchorcolor=blue
,citecolor=blue
,filecolor=blue
,linkcolor=blue
,menucolor=blue
,linktocpage=true
,pdfproducer=medialab
,pdfa=true
]{hyperref}
\usepackage[utf8]{inputenc}
\usepackage{url}

\newcolumntype{L}[1]{>{\raggedright\let\newline\\\arraybackslash\hspace{0pt}}m{#1}}
\newcolumntype{C}[1]{>{\centering\let\newline\\\arraybackslash\hspace{0pt}}m{#1}}
\newcolumntype{R}[1]{>{\raggedleft\let\newline\\\arraybackslash\hspace{0pt}}m{#1}}

\newcommand{\abs}[1]{\left\lvert #1 \right\rvert}

\newcommand {\be} {\begin {equation}}
\newcommand {\ee} {\end {equation}} 

\newcommand {\bes} {\begin {equation*}}
\newcommand {\ees} {\end {equation*}}

\newcommand{\beq}{\begin{equation}}
\newcommand{\eeq}{\end{equation}}
\newcommand{\bea}{\begin{eqnarray}}
\newcommand{\eea}{\end{eqnarray}}
\newcommand{\Eq}[1]{Eq.~(\ref{#1})}
\newcommand{\Eqs}[2]{Eqs.~(\ref{#1}) and (\ref{#2})}
\newcommand{\Sec}[1]{Sec.~\ref{#1}}

\newcommand{\Fig}[1]{Fig.~\ref{#1}}
\newcommand{\Figs}[2]{Figs.~\ref{#1} and \ref{#2}}
\renewcommand{\Ref}[1]{Ref.~\cite{#1}} 
\newcommand{\Refs}[1]{Refs.~\cite{#1}}
\newcommand{\Table}[1]{Table~\ref{#1}}

\newcommand{\Rsun}{R_{0}}

\newcommand{\msol}{M_{\odot}}
\newcommand{\pc}{\rm{pc}}
\newcommand{\msolpcsq}{\msol\ \pc^{-2}}
\newcommand{\msolpccu}{\msol\ \pc^{-3}}
\newcommand{\kpc}{\rm{kpc}}
\newcommand{\mssq}{\rm{m}\ \rm{s}^{-2}}
\newcommand{\mpl}{M_{\rm Pl}}
\newcommand{\abar}{\bar{a}}
\newcommand{\rhobar}{\rho_0} 
\newcommand{\azero}{a_0}
\newcommand{\grad}{\vec{\nabla}}
\newcommand{\gradphi}{\vec{\nabla}\phi}
\newcommand{\gradphisq}{(\vec{\nabla}\phi)^2}
\newcommand{\aphvec}{\vec{a}_{\rm ph}}
\newcommand{\abvec}{\vec{a}_{\rm b}}
\newcommand{\admvec}{\vec{a}_{\rm dm}}
\newcommand{\aph}{a_{\rm ph}}
\newcommand{\ab}{a_{\rm b}}

\newcommand{\rhob}{\rho_{\rm b}}
\newcommand{\rhosf}{\rho_{\rm SF}}

\newcommand{\vbar}{\overline{v}_z}
\usepackage{csvsimple}

\newcommand{\obs}{\text{obs}}

\begin{document}

\title{A Preference for Cold Dark Matter over Superfluid Dark Matter \\ in Local Milky Way Data}

\preprint{PUPT XXXX}

\author{Mariangela Lisanti}
\affiliation{Department of Physics, Princeton University, Princeton, NJ 08544}

\author{Matthew Moschella}
\affiliation{Department of Physics, Princeton University, Princeton, NJ 08544}

\author{Nadav Joseph Outmezguine}
\affiliation{School of Physics and Astronomy, Tel-Aviv University, Tel-Aviv 69978, Israel}

\author{Oren Slone}
\affiliation{Princeton Center for Theoretical Science, Princeton University, Princeton, NJ 08544}

\date{\today}

\begin{abstract}
There are many well-known correlations between dark matter and baryons that exist on galactic scales. These correlations can essentially be encompassed by a simple scaling relation between observed and baryonic accelerations, historically known as the Mass Discrepancy Acceleration Relation~(MDAR). The existence of such a relation has prompted many theories that attempt to explain the correlations by invoking additional fundamental forces on baryons.  The standard lore has been that a theory that reduces to the MDAR on galaxy scales but behaves like cold dark matter (CDM) on larger scales provides an excellent fit to data, since CDM is desirable on scales of clusters and above.  However, this statement should be revised in light of recent results showing that a fundamental force that reproduces the MDAR is challenged by local Milky Way dynamics and rotation curve data between 5--18~kpc. In this study, we test this claim on the example of Superfluid Dark Matter. We find that a standard CDM model is preferred over a static superfluid profile assuming a steady-state Galactic disk and discuss the robustness of this conclusion to disequilibrium effects. This preference is due to the fact that the superfluid model over-predicts vertical accelerations, even while reproducing galactic rotation curves.  Our results establish an important criterion that any dark matter model must satisfy within the Milky Way. 
\end{abstract}
\maketitle

\section{Introduction}
The existence of dark matter (DM) is well-supported by observations over many length scales. Although the hypothesis of a simple cold and collisionless dark particle is extremely successful at explaining large-scale structure evolution, observations on galactic scales suggest that DM may require more complex interactions~\cite{Bullock:2017xww,Tulin:2017ara}.
Most strikingly, observations across many galaxies point to a tight correlation between the total acceleration and the acceleration inferred from baryons alone. This correlation, historically known as the Mass Discrepancy Acceleration Relation (MDAR), suggests that accelerations are predicted solely by a galaxy's baryonic distribution, even when DM-dominated~\cite{Sanders1990,McGaugh:2004aw}.

A model that explains the MDAR can also explain a variety of additional observations, including the Baryonic Tully-Fisher Relation~\cite{1977A&A....54..661T, McGaugh:2000sr,Lelli:2015wst}, the Radial Acceleration Relation~\cite{McGaugh:2016leg,2000ApJ...533L..99M,2001ApJ...563..694V,2019ApJ...873..106D}, and possibly the diversity problem~\cite{deNaray:2009xj,Oman:2015xda,Kaplinghat:2019dhn}.  Although these observations could be explained within the collisionless cold DM (CDM) paradigm via baryonic feedback~\cite{Oman:2015xda,Sawala:2015cdf,Fattahi:2016nld,Navarro:2016bfs,2019arXiv191006345G} or through the introduction of DM self interactions~\cite{Tulin:2017ara,Kamada:2016euw,Ren:2018jpt,Kaplinghat:2019dhn}, it is tempting to search for theories that produce the MDAR at a more fundamental level. In this work, we address precisely this class of models and demonstrate that kinematic properties of the Milky Way~(MW) disk can strongly constrain such theories.

\begin{figure*}
  \centering
  \includegraphics[width=6.6in]{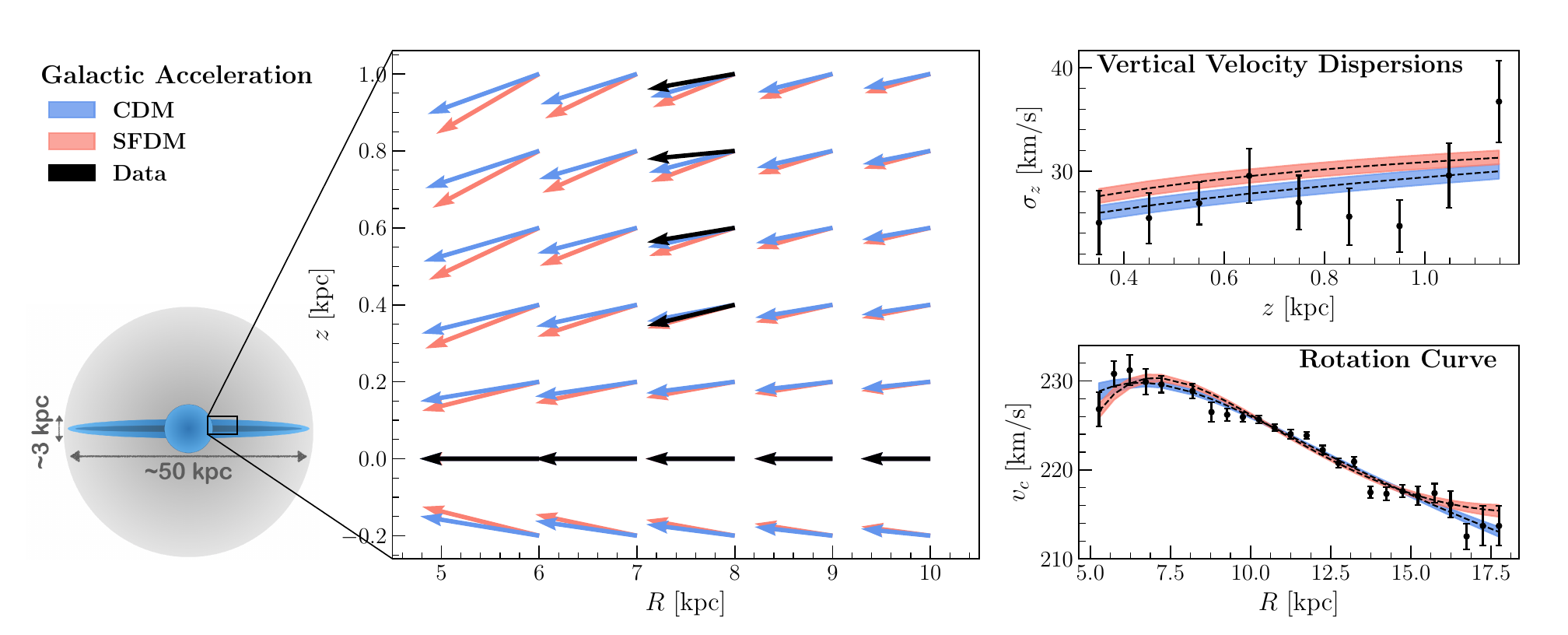}
  \caption{\textbf{Left Panel:}  Diagrammatic illustration of the source of tension for a model such as Superfluid Dark Matter (SFDM).  The arrows correspond to total accelerations for the best-fit CDM~(blue) or SFDM~(red) model, using the baryonic profiles obtained in this study.  The black arrows denote the expected accelerations as inferred from data~\cite{Zhang:2012rsb, Eilers:2019a}.  \textbf{Right Panel:}  The vertical velocity dispersions for the intermediate metallicity stellar tracer population from~\Ref{Zhang:2012rsb} as a function of height above the Galactic midplane \emph{(top)}, and the rotational velocity from~\Ref{Eilers:2019a} as a function of Galactocentric radius \emph{(bottom)}.  The colored bands denote the 68\% containment region around the median value of the posterior distribution, given by the dashed black line, for the SFDM and CDM models.  Although both models reproduce the rotation curve (radial acceleration), the SFDM model systematically over-predicts the vertical velocity dispersion (and therefore also vertical accelerations, see Eq.~\eqref{eq:sigmaz} below).  This tension underlines the primary reason that the SFDM model is strongly constrained by the data.  Note that the vertical velocity dispersions for the metal-rich and metal-poor tracer populations of \Ref{Zhang:2012rsb} are provided in \Fig{fig:tracers}.
}
  \label{fig:cartoon}
\end{figure*}

One class of models that explains the MDAR is modified Newtonian dynamics (MOND)~\cite{Milgrom:1983pn,Milgrom:1983zz,Milgrom:1983pn,Famaey:2011kh}. However, these models struggle to explain large-scale observations~\cite{Milgrom:2014usa}.  Alternatively, one can write a model that retains the strengths of CDM on large scales and of MOND-like dynamics on galactic scales~\cite{Blanchet:2006yt,Blanchet:2008fj,Zhao:2008rq,Bruneton:2008fk,Li:2009zzh,Ho:2010ca,Ho:2011xc,Ho:2012ar,Khoury:2014tka}.
In this study, we will focus our discussion on the example of Superfluid DM~(SFDM)~\cite{Khoury:2014tka,Berezhiani:2015bqa,Berezhiani:2017tth,Berezhiani:2019pzd}.
In this case, the DM is comprised of a light $\mathcal{O}(\text{eV})$ scalar particle that condenses to form a superfluid core in the center of galaxies when the virial temperatures are below some threshold. Within the superfluid core, the relevant degrees of freedom are those of phonons that are coupled to baryons and mediate a long-range emergent force, thereby giving rise to the MDAR.  At higher virial temperatures like those of clusters, the theory behaves as CDM. 

There are a number of significant differences between MOND-like models and SFDM. First, within MOND the prediction for measured accelerations depends on the choice of an arbitrary function which interpolates between the Newtonian regime and the deep-MOND regime. On the other hand, SFDM does not have such freedom and the parameters of the theory itself set the analogue of the interpolation function. Second, while MOND models only enhance baryonic accelerations, for SFDM there are two contributions to accelerations beyond that of baryons---an emergent phonon-mediated MOND-like force and the gravity of the SFDM itself.

It is commonly believed that models such as SFDM, which reproduce the MDAR on galactic scales while retaining CDM behavior on larger scales, provide an excellent fit to the data. However, this statement should be revised in light of the results of \Ref{Lisanti:2018qam}. There, it was shown that, in a model where the MDAR arises from a fundamental MOND-like force, it is challenging to simultaneously explain rotation curve and vertical velocity dispersion data in the Solar vicinity. The underlying reason for this is that the MW's distribution of stars and gas contributes too little gravity to explain the rotation curve but enough to explain vertical velocity dispersions. Thus, MW data inform us that a substantial amount of acceleration enhancement is required in the radial direction although essentially none is required in the vertical direction locally.

The left panel of \Fig{fig:cartoon} illustrates these points for the specific case of SFDM.  The arrows correspond to total accelerations predicted by the best-fit baryonic profiles found in this work together with the best-fit CDM~(blue) or SFDM~(red) halo model. The black arrows indicate the accelerations inferred from the data in \Ref{Zhang:2012rsb, Eilers:2019a}.  Because the phonon-mediated force dominates in the SFDM model, accelerations are too large in the direction perpendicular to the disk plane.  This figure is meant to be illustrative, but it does not account for the uncertainties on the measurements.  As will be discussed below, we perform a full Bayesian analysis to test the consistency of the SFDM with MW observables, marginalizing over uncertainties in the baryonic distribution as well as other model parameters. We confirm the intuition built up from \Fig{fig:cartoon} and demonstrate that SFDM is disfavored with respect to CDM because it over-predicts vertical velocity dispersions even while reproducing rotation curves. This basic tension should arise in other DM models that, like SFDM, seek to explain the MDAR by reducing down to a MOND-like force on galactic scales.  Our results thus establish an important criterion that any DM model must satisfy within the MW.

This analysis assumes that the superfluid profile is static, which is the current state of the art in modeling~\cite{Berezhiani:2017tth}. Self-consistently, we also treat the MW disk to be in steady-state equilibrium. We address the ramifications of this assumption, given current results from data~\cite{Widrow:2012wu, 2013MNRAS.436..101W, Gomez:2012rd, 2018Natur.561..360A, 2019MNRAS.482.1417B, Haines_2019, 2019MNRAS.485.3134L, 2019MNRAS.486.1167B, 2019MNRAS.490..797C}, in the Appendix. 

\section{Phenomenology of SFDM}
The SFDM model put forward in \Refs{Berezhiani:2015pia, Berezhiani:2015bqa,Berezhiani:2017tth} describes a scalar DM particle  that condenses into a superfluid core inside galaxies. Phonons inside this core mediate a long-range interaction between baryons that gives rise to an emergent force corresponding to an acceleration, $\vec{a}_{\rm ph}$. Thus, the total acceleration is the sum of this acceleration, the Newtonian acceleration produced by baryons only, $\vec{a}_{\rm b}$, and the Newtonian acceleration produced by the SFDM core, $\admvec$,
\be
\vec{a}_{\rm tot} = \aphvec + \abvec + \admvec  \, .
\label{eq:a_tot}
\ee
As will be discussed below, the contribution of $\admvec$ is often negligible. When this is the case, the result has the following asymptotic behavior,
\be
a_{\rm tot} \approx 
\begin{cases} 
      a_{\rm b} & a_{\rm b} \gg a_0 \\
      \sqrt{a_0 a_{\rm b}} & a_{\rm b} \ll a_0 \, ,
\end{cases}
\label{eq:a_MDAR}
\ee
where $a_0$ is an acceleration scale that arises from parameters of the theory.  This is precisely the behavior required by the MDAR. 

\Refs{Berezhiani:2015pia, Berezhiani:2015bqa,Berezhiani:2017tth} postulate that the Lagrangian of the low-energy effective theory takes on a specific form that induces the phonon interactions. Accounting for finite-temperature effects and assuming a static profile with vanishing fluid velocity, the Lagrangian is
\be
\mathcal{L} = \frac{2}{3}\Lambda(2m)^{3/2} X \sqrt{|X+\beta m\Phi|} - \alpha\frac{\Lambda}{\mpl}\phi\rhob \, ,
\label{eq:lagrangian}
\ee
where $m$ is the mass of the SFDM particle, $\alpha$ is a dimensionless coupling constant, $\Lambda$ is an energy scale, $\phi$ is the phonon field, $\rhob$ is the baryon matter density, $X = -m\Phi - \gradphisq/2m$, $M_{\rm Pl}$ is the reduced Planck mass,  and $\Phi$ is the Newtonian gravitational potential. Here, $\beta$ is a dimensionless constant that parametrizes finite-temperature effects; for concreteness as in \Ref{Berezhiani:2017tth}, we consider the case $\beta=2$.\footnote{There are additional ways of accounting for finite-temperature effects; see Ref.~\cite{Berezhiani:2015bqa} for examples.}

In \Eq{eq:lagrangian}, the chemical potential of the superfluid, $\mu$, has been implicitly included in the boundary condition on $\Phi$; specifically, $\Phi(r=0)=-\mu/m$.

The phonon-induced acceleration experienced by baryons, as inferred from the Lagrangian, is
\be
\aphvec = \frac{\alpha\Lambda}{\mpl}\gradphi \,.
\label{eq:phonon_accel_1}
\ee
The gravitational potential, $\Phi$, and accelerations obey
\be
\nabla^2\Phi = -\vec{\nabla}\cdot(\abvec + \admvec) = 4\pi G\left(\rhosf + \rhob \right) \, .
\label{eq:poisson}
\ee
The SFDM density, $\rhosf$, is obtained by differentiating \Eq{eq:lagrangian} with respect to $\Phi$,
\be
\rhosf = \frac{2}{3}\frac{\abar^2\mpl^2}{\azero}\frac{\left(\aph^2 - 6\abar^2\Phi\right)}{\sqrt{\aph^2 - 2\abar^2\Phi}} \,,
\label{eq:rhosf}
\ee
where $\abar\equiv~\alpha\Lambda m/\mpl$ and $\azero=\alpha^3\Lambda^2/\mpl$; the latter is the acceleration scale required by the MDAR, \emph{i.e.}, by \Eq{eq:a_MDAR}. Note that the nonzero chemical potential implies a boundary condition,
\be
\rhosf(r=0) \equiv \rhobar = 2^{3/2}\frac{\abar^3\mpl^2}{\azero}\sqrt{\frac{\mu}{m}} \,,
\label{eq:rhosf0}
\ee
assuming $\aph\to 0$ as $r\to 0$.

The equation of motion for the phonon field, $\phi$, now gives a field equation for $\aphvec$. Ignoring a ``pure curl'' term (see Appendix~\ref{sec:superfluid_review}) and using the Poisson equation, we get a relation between $\aphvec$ and $\abvec$,
\be
\frac{\aph^2 -\frac{2}{3}\abar^2\Phi}{\sqrt{\aph^2 - 2\abar^2\Phi}} \, \aphvec = \azero \abvec \, .
\label{eq:phonon_approx}
\ee
In the limit $\aph^2 \gg \bar{a}^2\Phi$, \Eq{eq:phonon_approx} simply reduces to the low-acceleration limit of \Eq{eq:a_MDAR}, $\aph^2 = \azero\ab$.

For this SFDM phenomenology to exist, a number of requirements must be met.  Specifically, the average separation of the SFDM particles within a galaxy must be much smaller than their deBroglie wavelength. Also, the SFDM particles must interact sufficiently frequently to thermalize.  In practice, these requirements translate to a superfluid phase in the inner region  of a galaxy and a particle phase in the outer region.  For a MW-mass galaxy where the measurable rotation curve is within the superfluid core, this requires a mass $m\lesssim 4\ \rm{eV}$ and self-interaction cross section $\sigma \gtrsim 2\times 10^{-34}\ \rm{cm}^2$~\cite{Berezhiani:2015pia, Berezhiani:2015bqa,Berezhiani:2017tth}.  The transition radius between the two regimes occurs at $\mathcal{O}(50)$~kpc. Note that \Eqs{eq:poisson}{eq:phonon_approx} are independent of $\sigma$ and depend on $m$ only through the parameters ($\abar$, $\azero$, $\rhobar$). As these are the only physical parameters that emerge from the model parameters ($m$, $\alpha$, $\Lambda$, $\mu$), our  analysis holds for any value of $m\lesssim 4$~eV.

\section{Methodology}
We perform a Bayesian likelihood analysis to test the consistency of the SFDM model with local MW observables and compare it to a model with a standard CDM halo.  In each case, we determine the expected acceleration field, marginalizing over all free parameters, including those of the baryonic distribution.    Comparing to MW observations, we recover the posterior distributions for the free parameters of the models.  The statistical approach of this study closely follows that of Ref.~\cite{Lisanti:2018qam}. 

For the SFDM scenario, the total acceleration is a sum of three contributions, as summarized in \Eq{eq:a_tot}. The baryonic acceleration, $\vec{a}_b$, is computed by assuming a density distribution for the stars and gas in the MW, which we model as a stellar bulge, stellar disk, and gas disk. The stellar bulge follows a Hernquist profile~\cite{Hernquist:1990be} with fixed scale radius of 600~pc,\footnote{Because the bulge scale radius is $\mathcal{O}(\kpc)$~\cite{Bland-Hawthorn:2016aaa} and we only consider observational constraints at Galactocentric radii beyond $R\gtrsim 5~\kpc$, the analysis is only weakly dependent on its value.}  
 while the stellar and gas disks each follow a double-exponential density profile.  We treat the mass of the bulge, the scale length of the stellar disk, as well as the scale heights and overall normalizations of both disks as free parameters. The scale length of the gas disk is fixed to twice that of the stellar disk~\cite{2008gady.book.....B,Bovy:2013raa}.  

Given a model of the baryonic distribution, one can then use Eqs.~(\ref{eq:phonon_accel_1})-(\ref{eq:phonon_approx}) to obtain the phonon-mediated acceleration $\vec{a}_{\rm ph}$, which depends on the free parameters $\rhobar$, $\abar$, and $\azero$ (see Appendix~\ref{sec:superfluid_review}). The DM acceleration, $\vec{a}_{\rm dm}$, for the SFDM model is obtained from the Poisson equation together with the solution to \Eq{eq:rhosf}.  For the CDM scenario, the acceleration is obtained assuming a spherical generalized Navarro-Frenk-White (NFW) density profile~\cite{Navarro:2008kc},\footnote{The results of both this and our previous study~\cite{Lisanti:2018qam} point towards a preference for any model that is able to enhance radial acceleration without enhancing vertical acceleration.  Therefore, we do not expect that slight variations to the DM profile will significantly affect our conclusions.  We have also explicitly verified that using a cored profile does not affect our results.} where the normalization, scale radius and inner slope are treated as free parameters.

We include several measurements of MW parameters in the likelihood to constrain the baryonic distribution. The scale length (scale height) of the stellar disk are taken to be $h_{*,R,\obs} = 2.6\pm0.5~\kpc$ ($h_{*,z,\obs} = 310\pm50~\pc$).  These values correspond to an \emph{effective} disk distribution for both the thin and thick disk contributions, whose respective properties we take from \Ref{Bland-Hawthorn:2016aaa}.  Because the scale height of the gas disk is not well-measured directly, we use the local density of interstellar gas at  the Galactic midplane $\rho_g(\Rsun,0) = 0.041\pm 0.004~\msol~\pc^{-3}$ from \Ref{McKee:2015hwa}.  We also use measurements of the local stellar ($j = *$) and gas ($j = g$) surface densities, defined as
\be
\Sigma_j^{z_\text{max}} = 2\int_0^{z_\text{max}}\ \rho_j(\Rsun,z')\ d z' \,,
\ee
adopting $\Sigma_{*,\obs}^{1.1} = 31.2\pm 1.6~\msolpcsq$ and $\Sigma_{g,\obs}^{1.1}=12.6\pm1.6~\msolpcsq$ at $z_\text{max}=1.1~\kpc$ as fiducial values~\cite{McKee:2015hwa}.  Measurements of the stellar bulge mass from photometry and microlensing exist but have a relatively large spread~\cite{Licquia:2014rsa}, so we conservatively ignore this constraint.

The radial dynamics of the MW are constrained by its rotation curve, which has recently been well-measured between 5 and 25 kpc from the Galactic Center using data from \textit{Gaia} Data Release 2~\cite{Eilers:2019a}.  We adopt the tabulated data from \Ref{Eilers:2019a} by symmetrizing the quoted statistical errors, giving circular velocities $v_{c,\obs}(R_i)$ at 26 different Galactocentric radii $R_i$ between 5 and 18 kpc from the Galactic Center.  We conservatively exclude measurements at larger $R$ where known systematic uncertainties may increase~\cite{Eilers:2019a}.

The vertical dynamics of the MW are constrained by measurements of the number densities and velocity dispersions of three mono-abundance stellar populations at $R=R_0$, which are provided in \Ref{Zhang:2012rsb}.
These results were obtained using 9000 K-dwarfs in the SEGUE sub-survey of the Sloan Digital Sky Survey (SDSS).  For each population, $i$, number densities, $n_{i,\obs}(z_k)$, and vertical velocity dispersions, $\sigma_{z,i,\obs}(z_k)$, were obtained for several values of $z_k$ between 300~pc and 1200~pc. Following \Ref{Zhang:2012rsb}, we model the number densities as
\be
n_i(z) = \tilde{n}_i\exp\left(-\frac{\abs{z}}{h_i}\right) \, ,
\label{eq:number_density}
\ee
where $\tilde{n}_i$ and $h_i$ are six additional free parameters.  The tracer stars are assumed to be in steady state and to be well-described by the Jeans equations. In this case, the vertical velocity dispersion is
\be
\sigma_{i,z}(z)^2 = \frac{-1}{n_i(z)}\int_z^\infty n_i(z')\, a_z(z')\, dz' \, ,
\label{eq:sigmaz}
\ee
where $a_z$ is the predicted vertical acceleration.

We constrain the parameters of each model using a Bayesian framework. The likelihood function is 
\be
\label{eq:likelihood}
\mathcal{L}(\boldsymbol{\theta}_\mathcal{M}) \propto \exp\left[-\frac{1}{2}\sum_{j}\left(\frac{X_{j,\obs}-X_j(\boldsymbol{\theta}_\mathcal{M})}{\delta X_{j,\obs}}\right)^2\right] \, ,
\ee
where $\boldsymbol{\theta}_\mathcal{M}$ is the set of free parameters of the model $\mathcal{M}=\rm{CDM},\rm{SFDM}$, $X_{j,\obs}\pm\delta X_{j,\obs}$ are the measured values and uncertainties of the observables used in this study, and $X_j(\boldsymbol{\theta}_\mathcal{M})$ are the values of the observables predicted by $\mathcal{M}$.  We use \texttt{MultiNest}~\cite{Feroz_2009,Buchner_2014} with \texttt{nlive}~$= 400$ to recover the posterior distributions for the model parameters as well as the Bayesian evidence for each model.  The specific priors used are provided in Table~\ref{tab:priors}.

\section{Results}
The main kinematic data that drive the model preference are shown in the right panel of \Fig{fig:cartoon}.  The top panel shows the vertical velocity dispersion data for the intermediate metallicity stellar tracer population from \Ref{Zhang:2012rsb}, while the bottom panel shows the MW rotation curve data from \Ref{Eilers:2019a}.  For each case, the dashed lines indicate the median of the posterior distribution for SFDM~(red) and CDM~(blue) models, with the colored bands representing the 68\% containment. 

Both the SFDM and CDM models reproduce the MW rotation curve data up to radial distances of $\sim 18$~kpc.  The SFDM model systematically over-predicts the rotational velocity towards the upper end of this range because the acceleration from the cored DM halo becomes comparable to the phonon acceleration.  The primary discrepancy between the kinematic predictions of the SFDM and CDM models is due to the vertical velocity dispersions near the Solar position.  The SFDM model clearly over-predicts measured dispersions.  The dominant contribution to the vertical acceleration for the CDM model comes from baryons, with only a subdominant contribution from the spherical DM halo.  In contrast, for the SFDM model, the phonon contribution dominates over the baryons even in the vertical direction. This results from the fact that the phonon-mediated force is parallel to the baryon-only gravitational force. 

\begin{figure}
  \centering
  \includegraphics[width=3.5in]{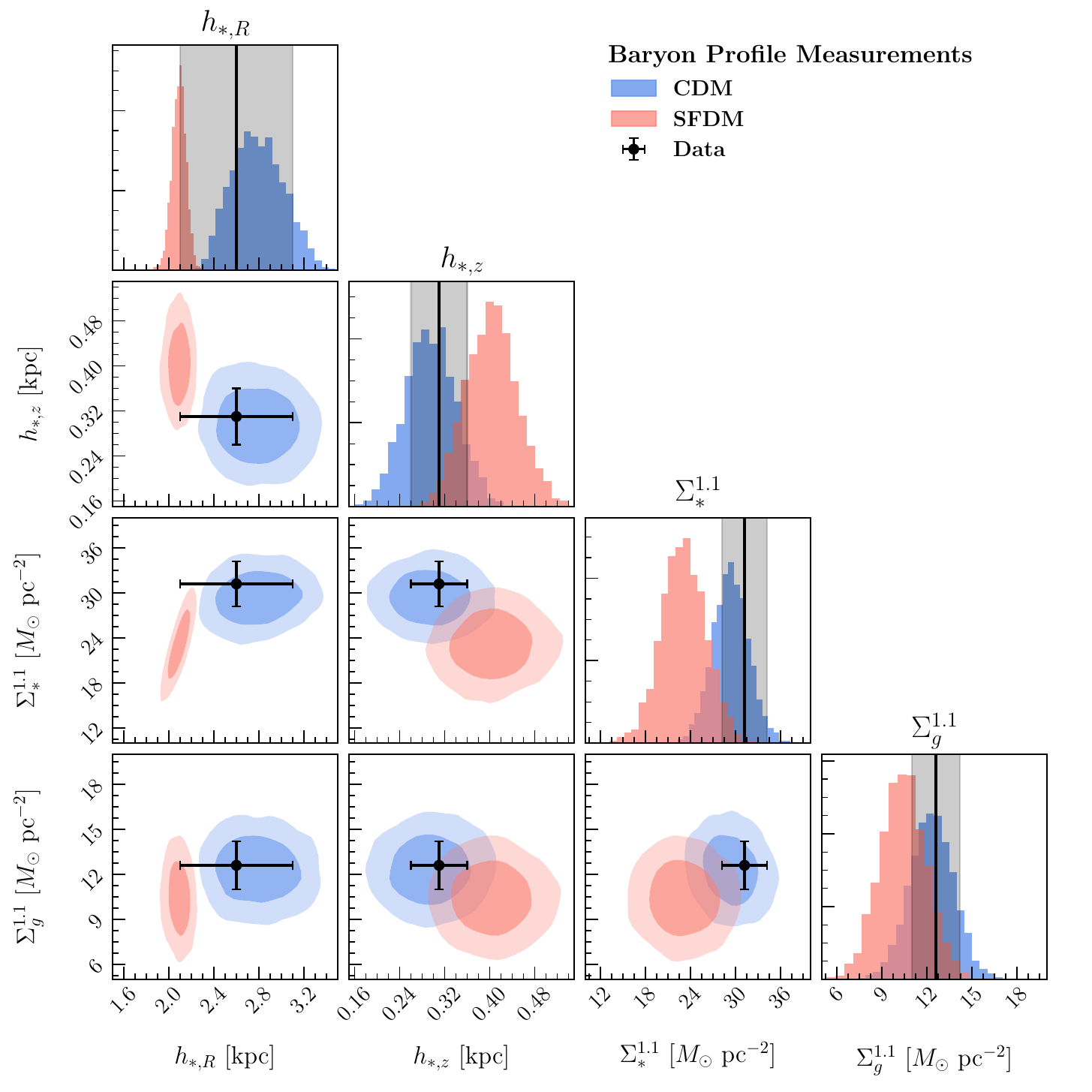}
  \caption{Posterior distributions of the stellar disk scale length and height ($h_{*, R}$ and $h_{*, z}$) and stellar and gas surface densities ($\Sigma_*^{1.1}$ and $\Sigma_g^{1.1}$) are provided for the SFDM (red) and CDM (blue) models. The measured values of each of these observables~\cite{Bland-Hawthorn:2016aaa, McKee:2015hwa} are indicated by the black points/lines along with their associated $1\sigma$ uncertainty.  The correlations for each pair of parameters are shown with the 68\% and 95\% regions of posterior probability indicated.  In general, the CDM halo results in a baryonic distribution that is more consistent with the MW observations.  The midplane gas density ($\rho_g(R_0, 0)$), which is accurately reproduced by both the SFDM and CDM models, is not shown here.     }
  \label{fig:baryon_measurements}
\end{figure} 

\Fig{fig:baryon_measurements} shows the posterior distributions for several  baryonic profile parameters for the SFDM~(red) and CDM~(blue) models.  Except for the midplane gas density, which both models accurately reproduce, there is typically a discrepancy between the SFDM model and the measured stellar and gas disk parameters.  The distributions for the CDM model are nicely consistent with the observed values, indicated by the black points/lines in the figure.  Relative to the CDM case, however, the SFDM model under-predicts the stellar scale length $h_{*,R}$, as well as the stellar and gas surface densities, $\Sigma_*^{1.1}$ and $\Sigma_g^{1.1}$, and over-predicts the stellar scale height, $h_{*,z}$.  These differences are driven by the fact that the SFDM scenario leads to vertical accelerations that are too large compared to observations, and thus the fitting procedure adjusts the baryonic profile as much as possible to reduce them.  

We find that the local MW data considered in this study strongly prefer the CDM model over the SFDM model with a Bayes factor~(BF) of $\ln \rm{BF}\approx32$.  We have verified that the BF is not dependent on small changes in our choice of priors and is driven primarily by the maximum likelihood of each model. We find that the chi-squared statistic, $\chi^2 = -2\ln \hat{\mathcal{L}}$, where $\hat{\mathcal{L}}$ is the maximum likelihood, is 77 (144) for the CDM (SFDM) model, with each model having 68 degrees of freedom. Additional plots relating to the models are provided in Appendix~\ref{sec:Accelerations}.


\section{Additional Cross-Checks}
We have performed the following additional tests of our analysis:  (1) repeating the likelihood analysis using each mono-abundance population from \Ref{Zhang:2012rsb} on its own, discarding the other two and (2) repeating the analysis for all three tracer populations, but with uncertainties on the velocity dispersion and number density measurements artificially increased by a factor of two. The latter test addresses the deviations in the data of \Ref{Zhang:2012rsb} from the smooth model prediction.\footnote{For this test, we find that $\chi^2$ for the CDM (SFDM) model is 44 (87) with 68 degrees of freedom. This suggests that the artificial increase in the uncertainties is more than enough to allow the CDM model, but not the SFDM model, to fully describe the data.} 
In both cases we find that the Bayes factor decreases but is always $\ln\rm{BF}\gtrsim 12$ and therefore does not change the qualitative conclusion that the CDM model is preferred over the SFDM model.

An additional concern is that the Jeans analysis might be altered by the observed disequilibrium of disk stars~\cite{Widrow:2012wu, 2013MNRAS.436..101W, Gomez:2012rd, 2018Natur.561..360A, 2019MNRAS.482.1417B, Haines_2019, 2019MNRAS.485.3134L, 2019MNRAS.486.1167B, 2019MNRAS.490..797C}.
A detailed discussion of these corrections is provided in Appendix~\ref{sec:diseq}, but the conclusions are summarized here.  In particular, we estimate that our results are robust to disequilibrium corrections to the vertical acceleration of at least $\sim 30\%$.   This is determined by performing additional analyses where the effect of disequilibrium is  artificially modeled as a constant reduction in the vertical acceleration.  Although this does not fully capture the effects of density waves in the Milky Way, it does represent a deviation from equilibrium that is maximally good for SFDM.  As such, it is a conservative test.
  We test reductions of 20, 30, and 40\%, finding that CDM is still preferred in the case of 20\% and 30\% reductions with Bayes factors of $\ln \mathrm{BF} \approx 13$ and $8$, respectively. For a 40\% reduction, SFDM is preferred with Bayes factor $\ln\mathrm{BF}\approx -14$; however, in this case, the $\chi^2$ statistic for the CDM (SFDM) model is 102 (87), with 68 degrees of freedom for both models. This suggests a poor goodness of fit for both scenarios. In this case, the minimum halo mass in the SFDM model is also too large to be consistent with observations and the value of $a_0\gtrsim 6\times 10^{-10}\ \mathrm{m}\ \mathrm{s}^{-2}$ (see \Figs{fig:all_params}{fig:min_halo}).
  
As a complementary probe of disequilibrium effects, we also consider restricting our analysis to $z$~values close to the midplane. Results in the literature~\cite{2017MNRAS.464.3775B,Haines_2019} suggest that disequilibrium corrections to the vertical acceleration decrease as one moves closer to the midplane. Accordingly, we repeat our analysis using only data within $700\ (600)\ \mathrm{pc}$ of the midplane and find that the Bayes factor decreases to $\ln\rm{BF}=9.2\ (1.7)$. Note that the integral in Eq.~\eqref{eq:sigmaz} is still performed above $z_{\rm max}$.
  
We note that the presence of disequilibrium could affect the assumption of a static SFDM profile. However, as long as the relaxation time of the SFDM is much smaller than that of the baryons, which is the standard assumption required for the model~\cite{Berezhiani:2017tth}, this should be a subdominant effect.

A final concern regards the assumption of a smooth baryonic profile.  In~\Ref{Lisanti:2018qam}, we showed that known perturbations of the Milky Way's baryonic distribution cannot substantially affect the predictions of a MOND-like model within $1$~kpc of the Sun.  In this paper, we use rotation curve data out to larger distances, but verifying whether perturbations at these distances still affect our analysis is beyond the scope of this work.  If this turned out to be true, it could potentially affect our conclusions.


\section{Conclusions}
We have shown that local Milky Way data strongly prefer a standard spherical CDM halo over a static SFDM model, and we have discussed the robustness of our results to disequilibrium effects.  This preference is due to the fact that the emergent baryonic force is approximately parallel to the gravitational force from baryons only in the SFDM scenario.  This typically enhances the vertical acceleration beyond what is consistent with stellar data near the Solar vicinity.  In addition, the necessity of reducing the predicted vertical accelerations drives the likelihood fit to baryonic distributions that are clearly inconsistent with observations of the stellar and gas disk distributions.
To explain both large-scale dynamics as well as the MDAR, there has been substantial motivation to write down models that behave as particle DM on large scales, but result in a long-range force between baryons on smaller scales, mimicking the phenomenology of MOND~\cite{Blanchet:2006yt,Blanchet:2008fj,Zhao:2008rq,Bruneton:2008fk,Li:2009zzh,Ho:2010ca,Ho:2011xc,Ho:2012ar,Khoury:2014tka, Khoury:2014tka,Berezhiani:2015bqa, Berezhiani:2017tth}.  SFDM is a representative example of this broader class of theories.  Importantly, our results establish a criterion for any DM model, especially one that purports to reproduce the MDAR.  Namely, the enhancement in the vertical acceleration must be suppressed compared to that in the radial direction.  Otherwise, the model will either over-predict vertical accelerations, under-predict radial accelerations, and/or be inconsistent with measurements of the baryonic profile.  We therefore see that the local dynamics of the MW disk can provide relevant constraints on the fundamental properties of DM.

\begin{acknowledgments}
We thank J.~Khoury for integral feedback over the course of this work.  We also thank M.~Geller, M.~Milgrom, S.~McGaugh, H.~Verlinde, and T.~Volansky for useful conversations. 
ML is supported by the DOE under contract DESC0007968 and the Cottrell Scholar Program through the Research Corporation for Science Advancement.  MM is supported by the DOE under contract DESC0007968. NJO is supported by the Azrieli Foundation Fellows program.  The work presented in this
paper was performed on computational resources managed and supported by Princeton Research Computing,
a consortium of groups including the Princeton Institute
for Computational Science and Engineering (PICSciE)
and the Office of Information Technology’s High Performance Computing Center and Visualization Laboratory
at Princeton University. This research made use of the \texttt{Astropy}~\cite{2013A&A...558A..33A}, \texttt{IPython}~\cite{PER-GRA:2007}, \texttt{matplotlib}~\cite{Hunter:2007}, \texttt{numpy}~\cite{Oliphant:2015:GN:2886196}, \texttt{galpy}~\cite{2015ApJS..216...29B}, \texttt{corner}~\cite{Foreman-Mackey:corner}, and \texttt{MultiNest}~\cite{Feroz_2009,Buchner_2014} software packages. 
\end{acknowledgments}
\twocolumngrid

\begin{appendix}

\setcounter{equation}{0}
\setcounter{figure}{0}
\setcounter{table}{0}
\setcounter{section}{0}
\makeatletter
\renewcommand{\theequation}{A\arabic{equation}}
\renewcommand{\thefigure}{A\arabic{figure}}
\renewcommand{\thetable}{A\arabic{table}}

\section{Superfluid Dark Matter Model}
\label{sec:superfluid_review}

\subsection{Numerical Procedure} 
\label{sub:numerical_procedure}

The equation of motion for the phonon field is not strictly \Eq{eq:phonon_approx}, but rather
\be
\grad\cdot\left(\frac{\aph^2 - \frac{2}{3}\abar^2\Phi}{\sqrt{\aph^2 - 2\abar^2\Phi}}\aphvec\right) = \frac{\azero}{2\mpl^2}\rhob \, .
\label{eq:phonon_full}
\ee
Using the Poisson equation for baryons only,
\be
-\vec{\nabla}\cdot\abvec = 4\pi G \rhob \, ,
\ee
one recovers \Eq{eq:phonon_approx} under the assumption that there is no ``pure curl'' term which vanishes under the divergence operator. Assuming this to be a good approximation (this will be justified below) one can solve for the predictions of the SFDM model. Specifically, the Poisson equation, \Eq{eq:poisson}, and the phonon equation, \Eq{eq:phonon_approx}, must be simultaneously evaluated. These constitute a set of coupled, multi-dimensional, non-linear, partial differential equations. For computational tractability, we follow \Ref{Berezhiani:2017tth} and adopt a slight variation of their algorithm to obtain an approximate superfluid solution. For a set of specified model parameters as well as specified $\rhob$, $\abvec$, we proceed as follows:

\begin{enumerate}
\item We integrate \Eq{eq:phonon_full} and ignore the ``pure curl'' term, giving \Eq{eq:phonon_approx}.

\item We algebraically solve \Eq{eq:phonon_approx} for $\aphvec$ in terms of $\Phi$ and $\abvec$.

\item We substitute the algebraic solution for $\aph^2$ into \Eq{eq:rhosf}, making \Eq{eq:poisson} a non-linear equation for $\Phi$.

\item We solve \Eq{eq:poisson} for $\Phi$, replacing $\rhob$ (and by extension $\abvec$) with a spherical density profile, $\tilde{\rho}_{\rm b}$, that gives the same acceleration $\abvec(r)$ in the midplane of the disk as that of the cylindrical profile. The spherical approximation makes \Eq{eq:poisson} computationally feasible to solve. We denote this approximate solution for the potential as $\tilde{\Phi}$.

\item We use the spherical $\tilde{\rho}_{\rm b}$ and $\tilde{\Phi}$ to obtain a solution for the SFDM density profile $\rho_{\rm SF}$ via \Eq{eq:rhosf}.

\item We plug the result for $\rho_{\rm SF}$ and the full cylindrical $\rhob$ back into \Eq{eq:poisson} to obtain a better approximation for $\Phi$ (a step that was not previously done in \Ref{Berezhiani:2017tth}). The solutions for $\admvec$ and $\abvec$ are found using this equation as well.

\end{enumerate}

The quasi-spherical approximation used in this procedure is reasonable because slight variations in $\rho_\text{SF}$ have negligible effects on the total acceleration.

\subsection{Neglecting the Pure Curl Term} 
\label{sub:neglecting_the_pure_curl_term}

It remains for us to justify the approximation of a negligible ``pure curl" term. Without any approximations, \Eq{eq:phonon_approx} should take the form,
\begin{equation}
  \frac{a_{\mathrm{ph}}^{2}-\frac{2}{3} \bar{a}^{2} \Phi}{\sqrt{a_{\mathrm{ph}}^{2}-2 \bar{a}^{2} \Phi}} \vec{a}_{\mathrm{ph}}=a_{0} \left(\vec{a}_{\mathrm{b}}+\vec{S}\right) \, ,
\end{equation}
where $\vec{S}$ is a divergenceless vector field, \textit{i.e.}, a ``pure curl" term. Previously, \Ref{Brada:1994pk} showed that in cases when the magnitude of the baryonic acceleration is dictated by the baryonic potential, \textit{i.e.}, $a_{\rm b}(\vec{r})=a_{\rm b}\left(\Phi(\vec{r})\right)$, then $\vec{S}=0$ and \Eq{eq:phonon_full} reduces exactly to Eq.~\eqref{eq:phonon_approx}. However, for a general baryonic profile this need not be precisely correct.

In order to evaluate the accuracy of the assumption $\vec{S}=0$, we have numerically solved \Eq{eq:phonon_full}.  We rewrite \Eq{eq:phonon_full} in the following form,
\begin{equation}\label{eq:MOND_like}
  \vec{\nabla}\cdot\left[\mu\left(\frac{a_{\rm ph}}{a_0},\xi\right)\vec{a}_{\rm ph}\right]=\vec{\nabla}\cdot\vec{a}_{\rm b} \, ,
\end{equation}
with the definitions,
\begin{equation}
  \mu(x,\xi) = \frac{x^{2}-{\xi}/{3}}{\sqrt{x^{2}-\xi}}\quad ; \quad \xi = 2\left(\frac{\bar{a}}{a_{0}}\right)^{2}\Phi \, .
\end{equation}
This equation is then solved using a trivial generalization of the grid relaxation algorithm presented in \Ref{Milgrom:1986ib}. Since the algorithm is computationally demanding, it is not feasible to run a scan where we solve the exact equation for every point in parameter space. For this reason, we have performed the SFDM scan following steps (1)--(6) of Sec.~\ref{sub:numerical_procedure}, and then solving the exact equation \eqref{eq:MOND_like} for the best-fit parameters in order to verify that $\vec{S}$ is negligible in the region relevant for our study.

We compare the exact and approximate solutions in the region of the MW relevant for this study, namely, close to the midplane between $5\lesssim R\lesssim20$~kpc and in the vertical direction around the Solar position between $0.3\lesssim|z|\lesssim 1.5$~kpc.
In this region, we find that 
\begin{equation}
  	\frac{\left|a^{\rm exact}_{i} - a^{\rm approx}_{i}\right|}{\left|a^{\rm exact}_{i}\right|}\lesssim 7\%\, 
\end{equation}
for both radial and vertical components, $i=R,z$. Here, $a^{\rm approx}_{i}$ is the result for the acceleration obtained from the procedure outlined in \Sec{sub:numerical_procedure}, and $a^{\rm exact}_{i}$ is obtained from the solution to \Eq{eq:phonon_full} with the same model parameters. We also verify that the derived vector field, $\vec{S}$, is indeed divergenceless. 

\section{Supplemental Figures}
\label{sec:Accelerations}

This section provides several supplemental figures that are referenced in the main body of the Letter. Each figure is discussed in detail in its associated caption. \vspace{10mm}

\makeatletter\onecolumngrid@push\makeatother

\begingroup
\squeezetable
\begin{table*}[t]
\centering
\begin{tabular}{C{1.8cm}C{1.8cm}C{1.8cm}|C{1.8cm}C{1.8cm}C{1.8cm}|C{1.8cm}C{1.8cm}C{1.8cm}}
  \Xhline{3\arrayrulewidth}

  \hline
  \multicolumn{3}{c|}{SFDM Model} & \multicolumn{3}{c|}{CDM Model} & \multicolumn{3}{c}{Stellar Tracer Populations} \\
  \hline
  Parameter & Prior & Unit & Parameter & Prior & Unit & Parameter & Prior & Unit \\ 
  \hline
  \centering
  $\abar$   & [200, 2000] & $10^{-10}\ \mssq$     & $\tilde{\rho}_{\rm DM}$ & [0, 10]      & $\msolpccu$ & $\tilde{n}_i$ &  [1, 100] & $10^{-3}\ \pc^{-3}$ \\
 $\azero$  & [0.1, 10] & $10^{-10}\ \mssq$     & $\gamma$                & [0, 10] & -- & $h_i$       &  [0.1, 1] & kpc \\
 $\rhobar$ & [0.1, 10] & $10^{-3}\ \msolpccu$  & $r_s$                   & [1, 100] & kpc &                 &           &                     \\

  \Xhline{3\arrayrulewidth}
  \multicolumn{3}{c|}{Stellar Disk} & \multicolumn{3}{c|}{Gas Disk} & \multicolumn{3}{c}{Stellar Bulge} \\
  \hline
  Parameter & Prior & Unit & Parameter & Prior & Unit & Parameter & Prior & Unit \\
  \hline
  $\tilde{\rho}_*$ & [0, 5]   & $\msolpccu$ & $\tilde{\rho}_g$  & [0, 5]   & $\msolpccu$  & $M_{*,\rm bulge}$ & [0, 2] & $10^{10}\ \msol$ \\
  $h_{*,z}$        & [0, 0.6] & kpc         & $h_{g,z}$         & [0, 0.6] & kpc          & $r_{*,\rm bulge}$ & --    & kpc                \\
  $h_{*,R}$        & [1, 5]   & kpc         & $h_{g,R}$         & -- & kpc        &          &              &                       \\

  \Xhline{3\arrayrulewidth}  

\end{tabular}
\caption{
  Model parameters used in the Bayesian likelihood analysis and the associated prior range for each.
  The SFDM model parameters $\abar$, $\azero$, and $\rhobar$ are defined in \Eqs{eq:rhosf}{eq:rhosf0}.
  The CDM model parameters correspond to the NFW profile $\rho_{\rm DM} = \tilde{\rho}_{\rm DM}(r/r_s)^{-\gamma}(1+r/r_s)^{\gamma-3}$, where $\gamma$ is the inner slope, $r_s$ is the scale radius, and $\tilde{\rho}_\text{DM}$ is the overall normalization.
  The parameters $\tilde{n}_i$ and $h_i$ model the number density of the stellar tracer populations used in this study, as defined in \Eq{eq:number_density}.
  The stellar and gas disks are modeled by double-exponential density profiles $\rho_{j,\rm disk} = \tilde{\rho}_j \exp\left(-R/h_{j,R}-|z|/h_{j,z}\right)$, where $h_{j,R}$ ($h_{j,z}$) is the scale length (height), $\tilde{\rho}_j$ is the overall normalization, and the subscript $j=*$ ($g$) corresponds to the stellar (gas) disk.
  The stellar bulge is modeled by a Hernquist density profile $\rho_{*,\rm bulge} = (M_{*,\rm bulge}/2\pi)\ (r/r_{*,\rm bulge})^{-1}(r+r_{*,\rm bulge})^{-3}$, where $M_{*,\rm bulge}$ is the mass and $r_{*,\rm bulge}$ is the scale radius of the stellar bulge.
  The model parameters that are considered free parameters in this study are shown with their associated prior range, where $[a,b]$ denotes a flat prior between $a$ and $b$.
  The parameters $h_{g,R}$  and $r_{*,\rm bulge}$ are not independent parameters in this study; specifically, $h_{g,R}$ is fixed to be $2h_{*,R}$ and $r_{*,\rm bulge}$ is held fixed at 0.6~kpc.
  }
\label{tab:priors}
\end{table*}
\endgroup

\begin{figure*}[h]
  \centering
  \includegraphics[width=7in]{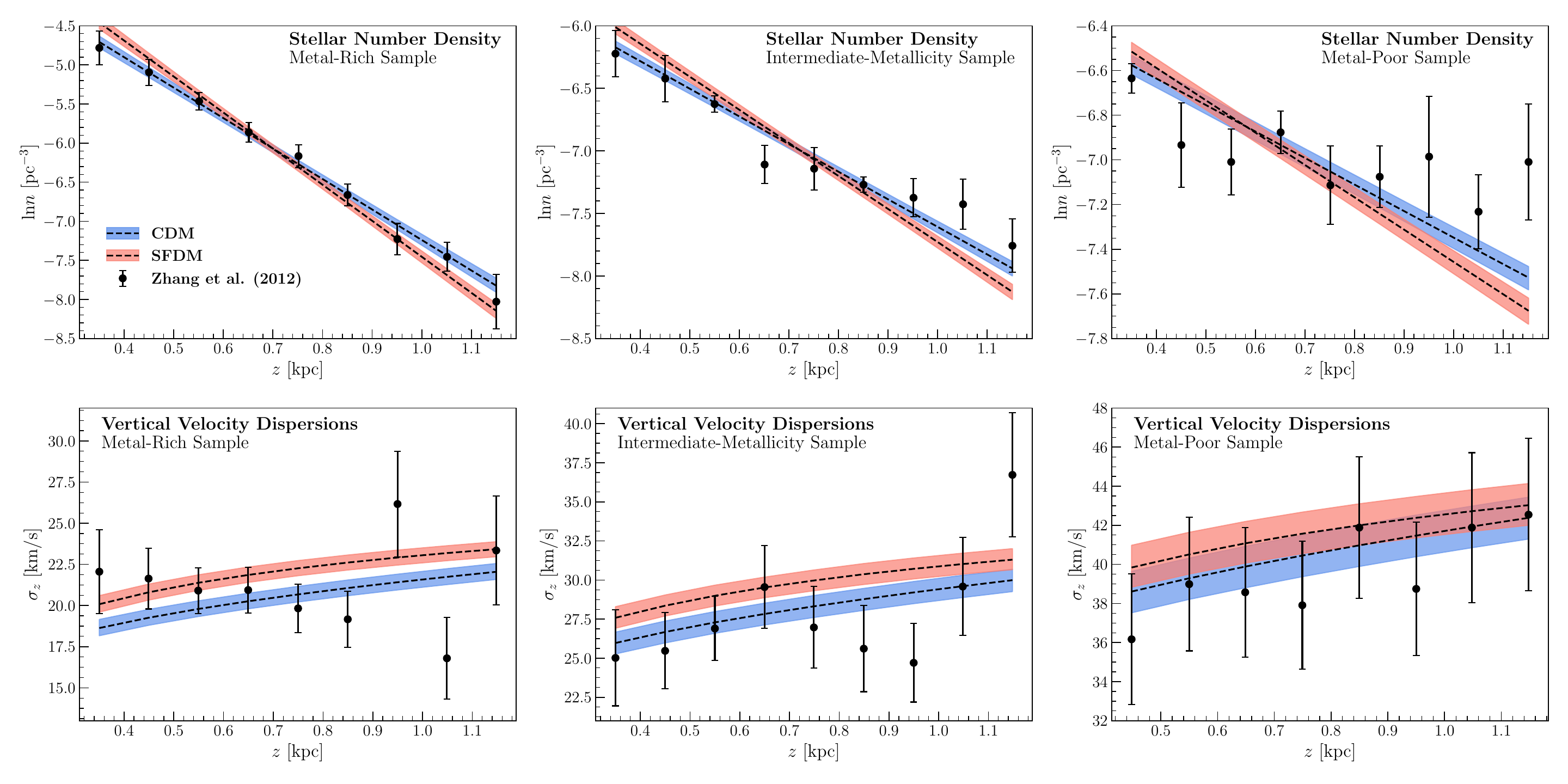}
  \caption{Number densities~\textbf{(top row)} and vertical velocity dispersions~\textbf{(bottom row)} for the three different stellar tracer populations used in this study. We refer to the populations by their metallicity:  metal-rich~\textbf{(left column)}, intermediate-metallicity~\textbf{(center column)} and metal-poor~\textbf{(right column)}.  The SEGUE data, indicated by the black points, are taken from \Ref{Zhang:2012rsb}.  The marginalized posterior distributions for the CDM~(blue) and SFDM~(red) models are also shown.  For each, the black dashed line corresponds to the median of the distribution, while the band is the 68\% containment region.  The SFDM model consistently over-predicts the vertical velocity dispersion for each tracer population.  This discrepancy is the primary driver for the significant tension we find for SFDM.}
  \label{fig:tracers}
\end{figure*}

\begin{figure*}[h]
  \centering
  \includegraphics[width=6in]{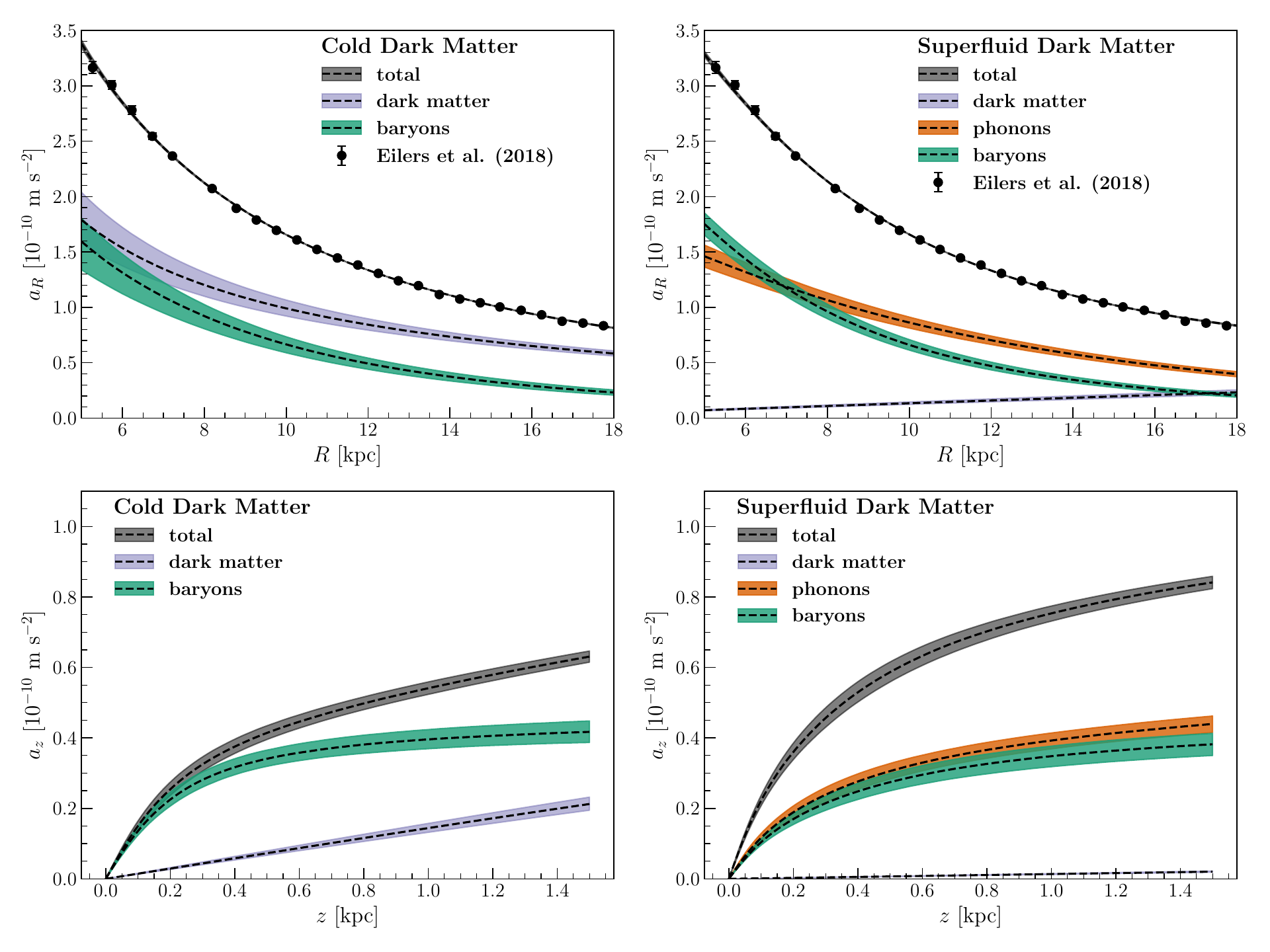}
\caption{Breakdown of contributions to radial and vertical accelerations for the CDM~\textbf{(left column)} and SFDM~\textbf{(right column)} models.  In each case,   
the total acceleration is shown in gray, and the contributions from the DM halo and baryons are shown in purple  and green, respectively.  For the SFDM model, the orange band corresponds to the phonon-induced acceleration.  The colored bands represent 68\% containment around the median value of the posterior distribution, given by the dashed black lines.  \textbf{Top row:}  Radial accelerations as a function of Galactocentric radius, $R$.  The original rotational velocity data and the associated uncertainties used in the likelihood (from \Ref{Eilers:2019a}) are given by black data points and corresponding error bars. In the radial direction, the phonon-mediated force contributes a dominant $\mathcal{O}(1)$ fraction in the SFDM model. Because of the large constant-density superfluid core in the SFDM model, the gravitational acceleration from DM increases linearly with radius and begins to dominate even over the contribution from phonons at large radii. \textbf{Bottom row:} Vertical accelerations as a function of height above the Galactic midplane, $z$.  In the vertical direction, the phonon-mediated force contributes a dominant $\mathcal{O}(1)$ fraction in the SFDM model. Since the phonon-mediated force enhances the vertical acceleration by the same factor as the radial acceleration, the SFDM model is unable to simultaneously reproduce the correct radial and vertical accelerations. In particular, because the SFDM model correctly fits the rotation curve, it over-predicts the vertical acceleration.  }
\label{fig:radial_accelerations}
\end{figure*}

\begin{figure*}[h]
  \centering
  \includegraphics[width=3.5in]{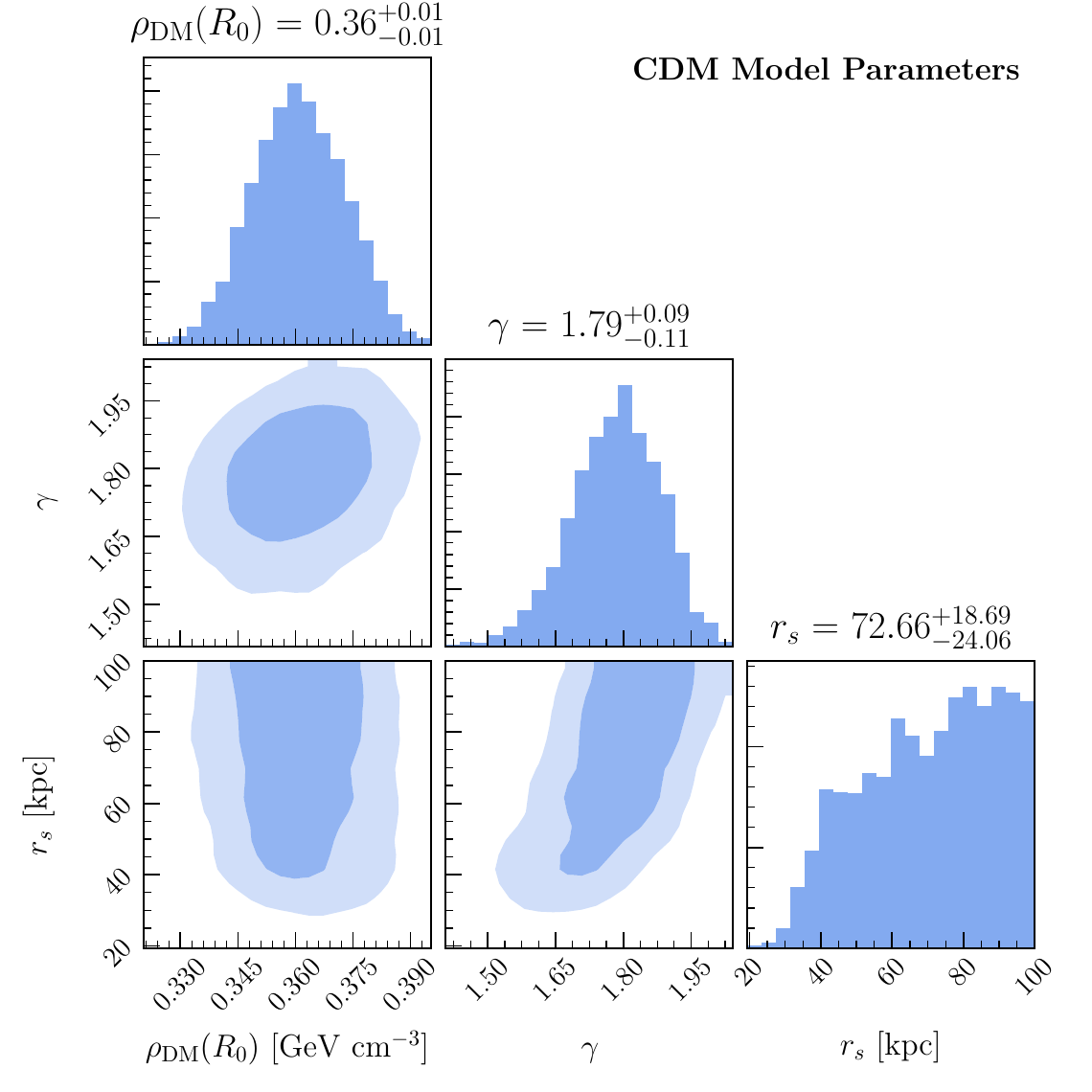}
    \includegraphics[width=3.5in]{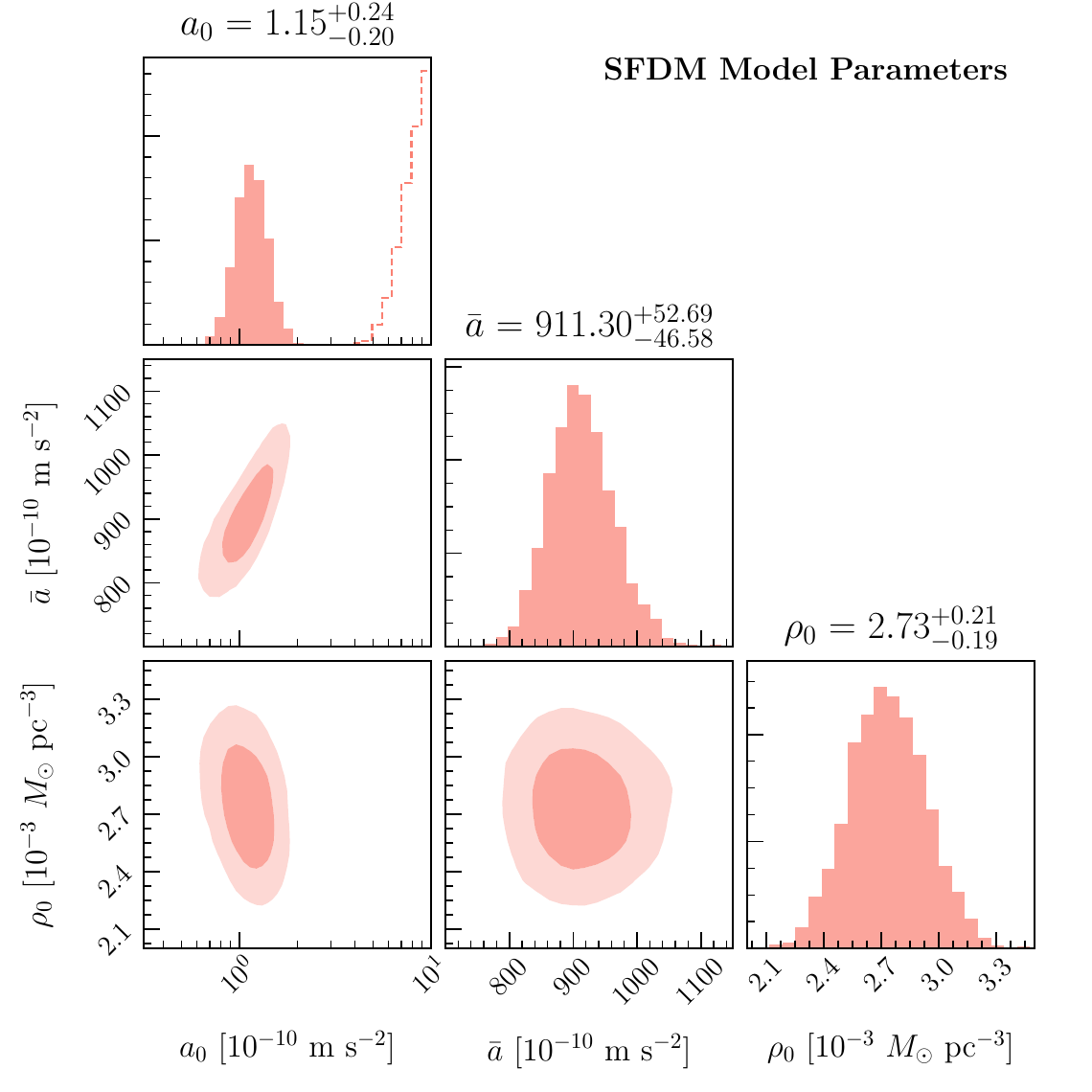}
    \caption{ Posterior distributions for the CDM \textbf{(left panel)} and SFDM \textbf{(right panel)} model parameters.  In the CDM case, we show the results for the inner slope, $\gamma$, scale radius, $r_s$, and local DM density, $\rho_\text{DM}(R_0)$.   In the SFDM case, we show the results for the accelerations $a_0$ and $\bar{a}$ and the superfluid density at the Galactic Center, $\rhobar$. The histograms show the distributions for each quantity, marginalized over all other fit parameters.  The two-dimensional correlations are also provided for each pairing; the shaded regions correspond to 68\% and 95\% contours.  Note that because $r_s$ is unconstrained by our analysis, it causes a steep inner slope with median $\gamma = 1.79$.  This is simply due to the fact that the fit is trying to reproduce an approximate power law $\propto r^{-2}$ over the range of radii considered to reproduce the flat rotation curve. Overlaid on the $a_0$ panel (dashed curve) is the posterior distribution obtained when including a $-40\%$ correction to the vertical acceleration (see Sec~\ref{sec:diseq} for details).}
  \label{fig:all_params}
\end{figure*}

\begin{figure*}[h]
  \centering
    \includegraphics[width=3.5in]{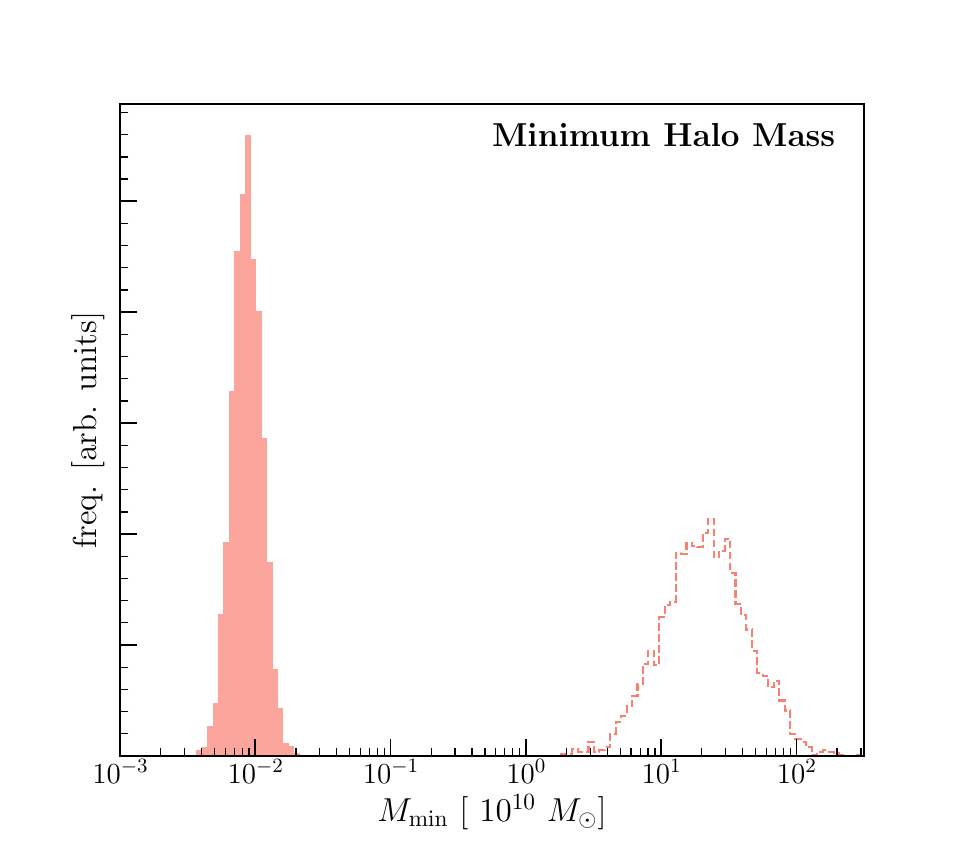}
  \caption{Posterior distribution for the minimum halo mass in the SFDM model.  This quantity is related to the SFDM parameters through the relation $M_\text{min} \simeq \Big( 12 \frac{\Lambda}{\text{meV}} \frac{m^3}{\text{eV}^3} \Big)^{-3}10^9~M_\odot$ --- see \Ref{Berezhiani:2017tth} for a derivation. Overlaid on this plot (dashed curve) is the posterior distribution obtained when including a $-40\%$ correction to the vertical acceleration (see \Sec{sec:diseq} for details).}
  \label{fig:min_halo}
\end{figure*}

\clearpage
\makeatletter\onecolumngrid@pop\makeatother

\section{Effects of Disequilibrium}
\label{sec:diseq}

The most recent stellar kinematic data from \textit{Gaia} show clear indications of disequilibrium behavior in the MW disk~\cite{Widrow:2012wu, 2013MNRAS.436..101W, Gomez:2012rd, 2018Natur.561..360A, 2019MNRAS.482.1417B, Haines_2019, 2019MNRAS.485.3134L, 2019MNRAS.486.1167B, 2019MNRAS.490..797C}. In particular, the observation of vertical density and velocity waves in stars of all ages~\cite{Widrow:2012wu,2019MNRAS.482.1417B} suggests that the MW disk is oscillating in response to a recent dynamical perturbation event. The assumption of steady-state equilibrium in this work is violated by the presence of time-dependent oscillatory perturbations in stellar density and velocity. Although a complete treatment of disequilibrium effects is beyond the scope of this work, it is important to consider the implications for the conclusions of this study. 

The effects of disequilibrium enter into our analysis through \Eq{eq:sigmaz}, where we have assumed steady-state equilibrium and ignored spatial inhomogeneities in the stellar distributions. There are two general ways that a perturbation event which violates these assumptions can affect the velocity dispersion through \Eq{eq:sigmaz}: by adding additional structure to the spatial dependence of quantities in \Eq{eq:sigmaz} and by adding an explicit time-dependent term into the equation.\footnote{We also note that the corrections to the vertical acceleration can differ above and below the mid-plane~\cite{Widrow:2012wu, 2013MNRAS.436..101W, Gomez:2012rd, 2018Natur.561..360A, 2019MNRAS.482.1417B, Haines_2019, 2019MNRAS.485.3134L, 2019MNRAS.486.1167B, 2019MNRAS.490..797C}, an effect that is not accounted for in our simple estimate.  Indeed, we are unable to test the implications of this as the data from~\Ref{Zhang:2012rsb} is only provided in terms of $|z|$.}

The first way that disequilibrium affects the spatial structure of quantities in \Eq{eq:sigmaz} is through the presence of spatial inhomogeneities. In particular, wave-like spatial variations in the density of stars in the MW disk have been observed in Refs.~\cite{Widrow:2012wu,2019MNRAS.482.1417B}. These spatial variations are not taken into account in our modeling of the MW disk gravitational potential nor of the number density of tracers in \Eq{eq:number_density}. However, the observed density inhomogeneities are reasonably well-modeled by a simple sinusoidal perturbation on top of a smooth distribution, allowing their effects to be estimated. Following \Refs{Widrow:2012wu,Read:2014qva}, one can write the total stellar density in the disk as $\rho(z) = \rho_{\rm eq}(z)(1+\delta(z))$ where $\rho_{\rm eq}(z)$ is the background exponential density profile and $\delta(z) \sim \delta_0 \sin kz$ with $\delta_0\sim 0.1$ and $k\sim 2\pi/(1\ \mathrm{kpc})$. The vertical force sourced by this density perturbation is approximately proportional to the surface density $a_z \propto \Sigma_z \sim 2\int_0^zdz'\ \rho(z')$. By carrying out this integral, \Refs{Widrow:2012wu,Read:2014qva} found that the density perturbations give a correction to the vertical gravitational force, $a_z$, smaller than $\delta a_z/a_z \lesssim 5\%$ at $z\sim 1\ \mathrm{kpc}$. As we will show below, such a correction is much smaller than the deviations from equilibrium behavior seen in detailed simulations; therefore, we do not further consider such modeling of the spatial inhomogeneities.

There is an additional effect of these spatial inhomogeneities on \Eq{eq:sigmaz} through the tracer densities $n_i(z)$, which are modeled as smooth equilibrium distributions in \Eq{eq:number_density}. Following the same treatment of \Refs{Widrow:2012wu,Read:2014qva}, one can write $n(z)=n_{\rm eq}(z)(1+\delta(z))$, where $\delta(z)$ can be taken to be the same as above (because the wave-like perturbations are known to be coherent in all stellar populations~\cite{2019MNRAS.482.1417B}), and compute the integral in \Eq{eq:sigmaz}. From this procedure, one finds that the effect of the spatial inhomogeneity through the tracer number density gives a correction to the velocity dispersion of $\delta \sigma_z/\sigma_z \lesssim 5\%$. We note that the relative uncertainty in the velocity dispersion measurements is $\sim 10\%$, so we do not consider this effect further.

The second way that disequilibrium can effect \Eq{eq:sigmaz} is through the addition of time dependence. In the absence of steady-state equilibrium, the time-derivative term in the collisionless Boltzmann equation must be included. To demonstrate this explicitly, the appropriate generalization of \Eq{eq:sigmaz}, which describes the vertical dispersion $\sigma_z$ for a given tracer, now includes a time-dependent term,
\be
\sigma_{z}^2(z) = \frac{-1}{n(z)}\int_z^\infty \left[n(z')\, a_z(z') + \frac{\partial \left( n \, \vbar \right)}{\partial t} \right] \, dz' \, .
\label{eq:sigmaz_dt}
\ee
Here, $n$ is the tracer number density, $a_z$ is the predicted vertical acceleration, and $\vbar$ is the average vertical velocity of the tracer population, where we have dropped the tracer index $i$ for simplicity.  
The time derivative in the integrand of \Eq{eq:sigmaz_dt} is the most dominant effect of disequilibrium; we note that it can be thought of as a correction to the vertical acceleration, $\delta a_z = \partial_t(n\vbar)/n$.  This is also the most difficult effect of disequilbrium to quantify because we do not know the precise time evolution of the perturbation event.  Numerical simulations suggest that the maximal correction to the vertical acceleration can be at least as large as $\delta a_z/a_z\sim 10\%$~\cite{2017MNRAS.464.3775B} or up to $\sim 50\%$~\cite{Haines_2019}.  The size and direction of the correction depend on the elapsed time since the event as well as the specific location in the disk, with under-dense regions being particularly affected.

To estimate the ramifications of the time-dependent term in \Eq{eq:sigmaz_dt} for the conclusions of our study, we repeated our analysis including an artificial, $z$-independent correction to the vertical acceleration. Specifically, inside the integral in \Eq{eq:sigmaz}, we take $a_z(z) = a_{z,\rm eq}(z)(1+\Delta)$, where $a_{z,\rm eq}$ is the unperturbed equilibrium acceleration and $\Delta$ is taken to be a fixed constant. This is an obvious oversimplification, as we know from, \emph{e.g.},~\Ref{Haines_2019} that the corrections can depend on the distance to the mid-plane.  We test the following three cases: $\Delta = - 0.2, -0.3,$ and $-0.4$.   Conservatively, we only consider corrections that decrease the vertical acceleration, as this works in favor of SFDM.  As discussed in the main text, SFDM over-predicts the vertical acceleration and thus a correction that decreases its overall value would bring it further inline with observations.

For $\Delta=-0.2~(-0.3)$, we find that $\ln\mathrm{BF}\approx 13\ (8)$, both of which still indicate strong statistical preference for the CDM model over the SFDM model. For the case of $\Delta=-0.4$, we find that $\ln\mathrm{BF}\approx -14$, corresponding to a strong preference for the SFDM model over the CDM model. This behavior is expected given that we are being conservative by considering only negative corrections to the acceleration which work in favor of the SFDM model. However, we find that although the SFDM model begins to show statistical preference over CDM (a decreased Bayes factor), it is forced into an unphysical range of parameter space. In particular, the SFDM model posteriors for both $\Delta=-0.3$ and $-0.4$ correspond to a minimum halo mass of $M_{\rm min}\gtrsim 5\times 10^{10}\ \mathrm{M}_{\odot}$, which is strongly in tension with known dwarf galaxies. Additionally, the posterior distribution for $a_0$ is peaked at an extremely large value (above $6\times 10^{-10}\ \mathrm{m}\ \mathrm{s}^{-2}$) and presses up against the prior of $a_0$ for both $\Delta=-0.3$ and $-0.4$. This prior was chosen to be consistent with measurements of the rotation curve of the MW at low radii~\cite{McGaugh:2019eic,Iocco:2015iia} and with the Radial Acceleration Relation~\cite{McGaugh:2016leg} in the deep-MOND regime.

Additionally, we test the goodness of fit by computing the chi-squared statistic. We find that for $\Delta=-0.2, -0.3$, and $-0.4$, $\chi^2 = 75~(105), 81~(107)$, and $102~(87)$, respectively, for the CDM (SFDM) models. We conclude that, for $\Delta=-0.2$ and $-0.3$, the qualitative conclusions of our analysis are unchanged, although the goodness of fit for the CDM model decreases in the case of $\Delta = -0.3$. In the case of $\Delta=-0.4$, although the SFDM model is favored over CDM, both models fail to fit the data well. 

Finally, we take a separate, but complementary, approach in testing the impact of disequilibrium on our baseline analysis.  As demonstrated in~\Ref{Haines_2019}, the corrections to the vertical acceleration tend to decrease towards the midplane.  Therefore, the time-dependent effects can be mitigated by choosing data as close to the midplane as possible, while maintaining differentiating power between the baryon and DM/phonon contributions to the vertical acceleration.  We thus investigate the robustness of the model comparison in this study to an artificial cutoff of the measured data. In \Table{tab:diseq}, we show the Bayes factor, representing the statistical preference for the CDM model over the SFDM model, for various values of $z_{\rm max}$, an artificial cutoff to the velocity dispersion and number density data. For a given value of $z_{\rm max}$, only the data measured at $z<z_{\rm max}$ (see \Fig{fig:tracers}) are used in the analysis.\footnote{Note that while the data for $\sigma_z$ and $n$ are not included above $z_{\rm max}$, the integration is still performed above this value. Therefore, disequilibrium effects above $z_{\rm max}$ could still potentially affect the parameter inference.}
As seen in \Table{tab:diseq}, as $z_{\rm max}$ becomes smaller, although the analysis gains in robustness to disequilibrium, there is decreased statistical power. This is to be expected because, by the arguments in the main body of this paper, the SFDM and CDM models are degenerate very close to the midplane. However, one can see that there is still substantial statistical preference for CDM at both $z_{\rm max}=600$ and $z_{\rm max} = (700)\ \mathrm{pc}$.



\begin{table}[ht]
\centering
\begin{tabular}{C{1.8cm}C{1.8cm}}
  \Xhline{3\arrayrulewidth}
  $z_{\rm max}$ [pc] & $\ln \rm{BF}$ \\
    \hline
  600 & 1.7  \\
  700 & 9.2 \\
  800 & 17 \\
  900 & 23 \\
  \Xhline{3\arrayrulewidth}
\end{tabular}
\caption{ The effect on the primary analysis of instituting an artificial cutoff on tracer number density and vertical velocity dispersion data. For a given value of $z_{\rm max}$, only data measured at $z<z_{\rm max}$ are used in the analysis. The quoted Bayes factor represents the statistical preference for the CDM model over the SFDM model. As the value of $z_{\rm max}$ becomes lower,  the statistical power of the dataset becomes weaker, while the robustness of the analysis to disequilibrium effects becomes greater.
  }
\label{tab:diseq}
\end{table}

\end{appendix}

\clearpage
\bibliographystyle{apsrev}
\bibliography{sf.bib}

\end{document}